\documentclass{article}
\usepackage{spconf,amsmath,graphicx,hyperref}

\usepackage{algorithm}
\usepackage{algorithmic}
\usepackage{amsmath}
\usepackage{amsfonts}
\usepackage{tabularx,booktabs}
\usepackage{multirow}
\usepackage{cite}

\usepackage{xcolor}

\title{Augmenting Large Audio Language Models with Low-Level Acoustic Features for Dysarthric Speech Detection}
\name{Mahdi Amiri$^{1 ,2}$, Hatef Otroshi Shahreza$^{1}$, Pascal Frossard$^{2}$, Ina Kodrasi$^{1}$ \thanks{This work was supportecd by the Swiss National Science Foundation project CRSII5\_202228 on ``Characterisation of motor speech disorders and processes''.}
}
\address{\textit{$^1$Idiap Research Institute, Switzerland} \\
\textit{$^2$École Polytechnique Fédérale de Lausanne (EPFL), Switzerland}\\}
\begin{document}
\ninept
\maketitle

\begin{abstract}
Automatic dysarthric speech detection approaches can support traditional clinical diagnosis, which relies on costly and time-consuming evaluation by a speech and language pathologist. Existing automatic approaches predominantly rely on deep learning (DL). More recently, Large Audio Language Models (LALMs) have emerged as a promising alternative given their strong performance across various tasks, but their application to dysarthric speech detection has not yet been established. We propose a framework that fine-tunes LALMs for dysarthric speech detection on speech recordings combined with textual information comprising low-level acoustic features and speaker demographics. Across two LALMs, our framework outperforms DL-based baselines, with Qwen2-Audio-Instruct achieving state-of-the-art performance. An ablation study shows that incorporating acoustic features and speaker demographics during fine-tuning improves LALM performance, while LALMs alone exhibit only chance-level zero-shot performance. These findings establish an effective approach for adapting LALMs to dysarthric speech detection.

\end{abstract}

\begin{keywords}
    dysarthric speech detection, Large Audio Language Models, supervised fine-tuning, low-level acoustic features
\end{keywords}  
    
\section{Introduction}
Neurological conditions such as Parkinson's disease or Amyotrophic frequently cause dysarthria, a speech impairment that can significantly hinder communication~\cite{duffy2012motor}. As the aging population grows, dysarthria is becoming increasingly common. Its early detection is critical, as it may serve as an early indicator of the underlying neurological disease \cite{duffy2012motor}. Traditionally, dysarthria has been diagnosed through time-consuming and costly auditory-perceptual assessments conducted by speech and language pathologists, motivating the development of automatic detection approaches to reduce the burden on the healthcare system. Earlier work has used traditional machine learning techniques~\cite{handcrafted_1} as well as Deep Learning (DL)–based approaches for automatic dysarthric speech detection~\cite{baseline, amiri2024test, amiri_suppress}.

Recently, with the rise of Large Audio Language Models (LALMs) and their promising performance on various speech-related tasks such as question answering, audio event detection \cite{ghosh2024gama, tangsalmonn}, or emotion recognition \cite{mai2025aa, shi2026plug, zhao2025steering},
researchers have started to explore the capabilities of LALMs for dysarthric speech detection. In \cite{amiri_incontext}, we proposed an in-context learning approach for this task. Although the approach achieved remarkable performance, the performance was lower than existing DL-based approaches.
In \cite{ortiz2026multimodal}, the authors find that LALMs are effective at estimating macro-descriptors of speech. They therefore adopt a two-step approach, using an LALM to extract these macro-descriptors in the first step and training a traditional classifier on the resulting descriptors in the second step. While effective, this approach still relies on a separate classifier rather than an LALM capable of performing detection directly. Further, their comparisons additionally rely on GPT-4o~\cite{hurst2024gpt}, whose license agreements do not permit use with datasets such as PC-GITA \cite{pc_gita}.
A separate line of work in \cite{mai2025aa} identifies an additional gap, \textit{i.e.}, LALMs rely primarily on semantic speech encoders such as the Whisper encoder \cite{radford2023robust}. Consequently, they largely neglect low-level acoustic features that are particularly important for tasks such as dysarthric speech detection.

These findings motivate our goal of developing an LALM that performs dysarthric speech detection directly, without a separate downstream classifier, while still making use of the low-level acoustic information that LALMs otherwise miss. To this end, we propose a framework that fine-tunes LALMs using speech recordings together with textual information comprising low-level acoustic features (\textit{e.g.}, loudness, speech rate, etc.) and speaker demographics. These acoustic features are extracted using the OpenSMILE framework \cite{eyben2010opensmile}. We consider the Qwen2-Audio-7B-Instruct \cite{qwen2} and Phi-4-multimodal-instruct \cite{abouelenin2025phi} models, both of which demonstrate strong performance in audio understanding across a wide range of downstream tasks~\cite{abouelenin2025phi}.
Our experiments demonstrate the effectiveness of incorporating low-level acoustic features and speaker demographics during fine-tuning, enabling both LALMs to achieve achieve state of the art performance than state-of-the-art DL-based approaches. Ablation studies show that while the models achieve competitive performance without the acoustic features and speaker demographics information, they achieve tate-of-the-art performance when this information is incorporated during fine-tuning. 

\begin{figure*}[t!]
  \centering
  \includegraphics[width=0.9\textwidth]{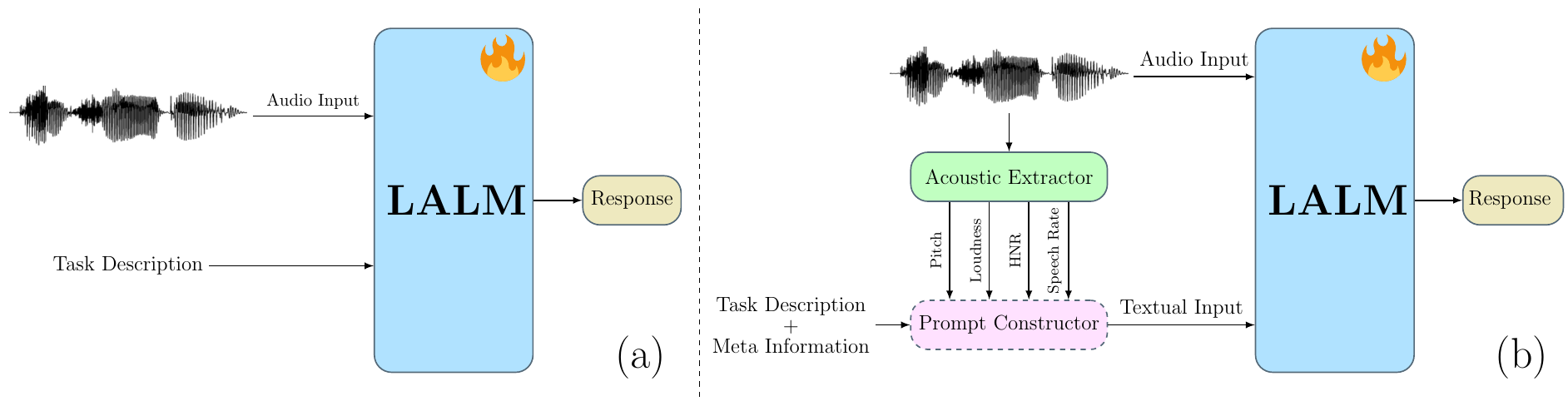}
  \caption{Schematic illustration of the proposed method: a) Conventional fine-tuning of the LALM using raw audio and textual inputs. b) Fine-tuning with the proposed approach, where besides the raw audio and textual inputs, low-level audio features are extracted and incorporated into the textual input together with the task description and metadata.}
  \label{fig:schematic}
\end{figure*}

The key contributions of this paper are as follow:

\begin{itemize}

    \item We propose a framework for fine-tuning LALMs for dysarthric speech detection by incorporating low-level acoustic features and speaker demographics as textual information alongside speech recordings.

    \item We conduct experiments using two LALMs and demonstrate that, when fine-tuned using our proposed framework, both models outperform state-of-the-art DL-based approaches for dysarthric speech detection.

    \item We conduct ablation studies to analyze the effect of incorporating low-level acoustic features and speaker demographics on the final performance, and to further evaluate the zero-shot capabilities of LALMs for dysarthric speech detection.
\end{itemize}

\section{Related Work}


\subsection{DL-based automatic dysarthric speech detection}
Traditional DL-based approaches for dysarthric speech detection typically combine time-frequency features with various architectures. Common features include short-time Fourier transform magnitude spectrograms~\cite{baseline} and Mel-frequency cepstral coefficients~\cite{mfcc_ref}, which are often combined with convolutional neural networks~\cite{baseline} or autoencoders~\cite{mel_ref_vae}. More recently, self-supervised foundation models such as wav2vec 2.0~\cite{w2v2}, XLS-R~\cite{xlsr}, and Whisper~\cite{radford2023robust} have instead been used to obtain representations directly. These representations are combined with classifiers such as multi-layer perceptrons for dysarthric speech detection~\cite{amiri2024adversarial, tilak}, outperforming time-frequency-based approaches and establishing the current state-of-the-art~\cite{amiri2024adversarial}. We consider these self-supervised approaches as our DL-based baselines.

\subsection{Large Audio Language Models}
LALMs achieve remarkable performance on many tasks such as audio understanding, captioning, and event detection \cite{tangsalmonn, ghosh2024gama}, and recent research has increasingly sought to leverage them for other speech-related tasks \cite{mai2025aa}. Despite this success across a wide range of downstream speech tasks, only a handful of studies have applied LALMs to dysarthric speech detection. In \cite{amiri_incontext}, we proposed an in-context learning approach for dysarthric speech detection, but the resulting model still lags behind DL-based baselines. In \cite{ortiz2026multimodal}, macro-descriptors are extracted with an LALM and used to train a separate classifier for Parkinson's disease detection, achieving strong results with GPT-4o Audio. However, this approach still relies on a downstream classifier rather than an LALM capable of detection on its own, and its best results depend on a closed-source model whose license agreements do not permit use with datasets such as PC-GITA \cite{pc_gita}. Together, these limitations motivate the development of an open-source LALM capable of performing dysarthric speech detection directly.


\section {Problem Formulation}
Let $\tilde{x}$ be a raw text input and $\tilde{a}$ be a raw audio waveform.
Let $\tau : \tilde{x} \mapsto \mathbf{x}$ be a tokenizer that maps the raw text
to a sequence of discrete tokens $\mathbf{x} = (x_1, x_2, \ldots, x_N)$, where
$N$ is the number of text tokens and $x_i \in \mathcal{V}$ for all $i$, with
$\mathcal{V}$ being the text vocabulary of size $|\mathcal{V}|$.

Let $\mathrm{enc}_a : \tilde{a} \mapsto \mathbf{a}$ be a learned audio encoder
with parameters $\phi$ that maps the raw waveform to a sequence of continuous
representations $\mathbf{a} = (a_1, a_2, \ldots, a_M) \in \mathbb{R}^{M \times d}$,
where $M$ is the number of audio frames and $d$ is the encoder hidden dimension. These representations are then passed through a projection layer $\mathrm{proj}_a$, parametrized b $\psi$, that maps them into the embedding space of the language model.
An \textit{audio language model} with parameters $\theta$ is a probabilistic model
that factorizes the conditional distribution over an output token sequence
$\mathbf{y} = (y_1, y_2, \ldots, y_T)$, where $T$ is the output length and
$y_i \in \mathcal{V}$ for all $i$, autoregressively as

\begin{equation}
\begin{split}
P_{\theta}(\mathbf{y} \mid \tau(\tilde{x}),\,  \mathrm{proj}_{\psi}(\mathrm{enc}_{\phi}(\tilde{a})))
        &= \\ \prod_{i=1}^{T} P_{\theta}\!\left(y_i \mid \tau(\tilde{x}),\right.
        &\left. \mathrm{proj}_{\psi}(\mathrm{enc}_{\phi}(\tilde{a})),\, y_{<i}\right),
\end{split}
\end{equation}
\noindent where $y_{<i} = (y_1, \ldots, y_{i-1})$ denotes all previously generated
tokens and each factor $P_{\theta}(y_i \mid \tau(\tilde{x}),  \mathrm{proj}_{\psi}(\mathrm{enc}_{\phi}(\tilde{a})),
y_{<i})$ is a categorical distribution over $\mathcal{V}$ parametrized by $\theta$.


Given a raw audio waveform $\tilde{a}$ and a textual input $\tilde{x}$, our objective is to fine-tune an LALM to perform automatic dysarthric speech detection. Specifically, given an input pair $(\tilde{a}, \tilde{x})$, the model aims to predict a label $y \in \{0,1\}$, where $y=1$ indicates dysarthric speech and $y=0$ indicates neurotypical speech. Formally, given an LALM parameterized by $\theta$, the prediction can be expressed as

\begin{equation}
    \hat{y} = f_{\theta}(\tilde{a}, \tilde{x}),
\end{equation}
where $f_{\theta}$ represents the LALM and $\hat{y}$ denotes the predicted class. The textual input $\tilde{x}$ contains the task instruction that guides the model to determine whether the given speech sample is dysarthric or neurotypical.

\begin{table*}[ht!]
\caption{Input prompts for different training strategies. For standard SFT, we input only the audio alongside the task description. For the feature-augmented method, we additionally provide the speaker's demographic information and low-level acoustic features alongside the task description and audio.}
\centering
\footnotesize
\begin{tabular}{|p{\textwidth}|}
\hline
\rule{0pt}{3ex}%
\textit{\textbf{Simple Training Prompt}}: \\
You are an AI agent trained to classify audio recordings into two classes. The two classes represent healthy individuals (Class 0) and individuals with dysarthria (Class 1).
The audio recordings may vary in duration. Your task is to analyze each recording and classify it into one of the two classes (0 or 1). Always provide a prediction, even if the audio is difficult to interpret. \\
\textless audio sample\textgreater\\
Please classify this recording into one of the two classes. 
\\[1ex]
\hline
\rule{0pt}{3ex}%
\textit{\textbf{Feature-augmented prompt}}:\\
You are an AI agent trained to classify audio recordings into two classes. The two classes represent healthy individuals (Class 0) and individuals with dysarthria (Class 1). Individuals with dysarthria may exhibit abnormalities in speech production such as monopitch, monoloudness, breathiness, or irregular speech rate. The audio recordings may vary in duration. Your task is to analyze each recording and classify it into one of the two classes (0 or 1). Always provide a prediction, even if the audio is difficult to interpret. \\
\textless audio sample\textgreater \\
This is a recording from a \textless gender \textgreater speaker speaking \textless language \textgreater. Please classify this recording into one of the two classes. \\
You are allowed to use the provided low-level acoustic features of the audio recording as supplementary information in your reasoning.
The mean pitch level is \textless xx\textgreater Hz. The standard deviation of the pitch is \textless xx\textgreater Hz. 
The mean perceptual loudness level is \textless xx\textgreater dB. 
The standard deviation of the perceptual loudness is \textless xx\textgreater dB. The mean harmonic-to-noise ratio is \textless xx\textgreater dB. The mean speech rate is \textless xx\textgreater syllable per second. \\
\hline
\end{tabular}
\vspace{-0.3cm}
\label{tab:prompt}
\end{table*}

\section{Proposed Method}
\label{sec:proposed}
An overview of our proposed method is shown in Fig.~\ref{fig:schematic}. A straightforward approach to adapting a pretrained LALM to our task is to fine-tune it using the audio sample and a textual task prompt. We use supervised fine-tuning (SFT) to classify speech as either neurotypical or dysarthric. The training objective is the standard cross-entropy loss computed only over the response tokens, with the prompt tokens masked from the loss. The objective is defined as
\begin{equation} \begin{aligned} \mathcal{L}_{\text{SFT}}(\theta) = - \frac{1}{|\mathbf{y}|} \sum_{t=1}^{|\mathbf{y}|} \log P_{\theta}\!\left(y_i \mid \tau(\tilde{x}),\right. \left. \mathrm{enc}_{\phi}(\tilde{a}),\, y_{<i}\right). \end{aligned} \label{eq:sft_loss} \end{equation}
Although this formulation is effective~(cf. Section~\ref{sec:exp_result}), performance can be further improved by incorporating additional features into the input. In \cite{mai2025aa} it is noted that most LALMs rely on semantic encoders and pay limited attention to acoustic information. Such information is particulatly important for dysarthric speech detection, since dysarthria is known to manifest in characteristics such as monopitch, monoloudness, and breathiness~\cite{duffy2012motor}. Moreover, incorporating speaker demographic information, such as sex and language, can further improve model performance. 

 Motivated by this, we propose a feature-augmented framework that adds a small set of acoustic descriptors and speaker demographics to the textual prompt. To this end, we compute the mean and standard deviation of pitch and loudness, the mean harmonic-to-noise ratio, and the speech rate. These descriptors were chosen because they are commonly associated with dysarthric speech characteristics in the clinical literature, rather than through any feature-selection or optimization procedure. 
 We view this choice as an illustrative starting point, and our framework readily accommodates other or additional descriptors. We include the raw feature values directly in the textual prompt rather than converting them into categorical labels as in \cite{mai2025aa}, and combine them with the speaker's demographic information. These additional information is then provided to the model alongside the raw audio for fine-tuning. An overview of the prompting strategies for both approaches is provided in Table~\ref{tab:prompt}. Our experimental results demonstrate that incorporating this information improves performance, enabling the LALMs to outperform the baseline methods while using only limited data for fine-tuning.


\section{Experimental Settings}


\hspace{\parindent}\textbf{Database.} \enspace
\textit{PC-GITA} \cite{pc_gita} is a Spanish dataset 
containing recordings from 50 patients with Parkinson’s disease (25 male and 25 female) and 50 neurotypical speakers (25 male and 25 female). The recordings were originally sampled at 44.1 kHz and were subsequently resampled to 16 kHz for our experiments. We use the monologue, read-text, and sentence subsets of PC-GITA dataset. 



\textbf{Pre-processing.} \enspace
We segment utterances into 30-second chunks with a 50\% overlap. The final residual segment of each utterance is also retained. 
We adopt the standard 10-fold cross validation framework as in \cite{amiri_suppress}, which reduces the influence of custom data splits and enables fair performance comparisons. To account for the randomness introduced during fold generation, we repeat the training using three different random seeds and report the mean and standard deviation across these runs.


\textbf{LALMs Architecture.} \enspace
We consider the following two LALMs.

\begin{itemize}
    \item \textit{Qwen2-Audio-7B-Instruct} \cite{qwen2} is a large-scale audio-language model capable of accepting audio input and performing audio analysis. Its audio encoder is based on the Whisper-large-v3 model \cite{radford2023robust}, and it uses Qwen-7B \cite{bai2023qwen} as its LLM. The model resamples the data to 16 kHz and converts it into a 128-channel mel-spectrogram using a 25 ms window size and a 10 ms hop size. The model performs best on audio under 30 seconds. For readability, we will refer to this model as Qwen2-Audio in the remainder of the paper.

\item \textit{Phi-4-multimodal-instruct} is a multimodal model that combines various modalities such as vision and speech \cite{abouelenin2025phi}. This model uses a \textit{"mixture of LoRAs"} architecture, in which the model employs modality-specific Low-Rank Adaptations (LoRAs). This allows the model to keep the language model frozen. The speech LoRA has a rank of 320. The input is transformed into 80-dimensional log-Mel filter-bank features with 10 ms frames. Unlike Qwen2-Audio, which has a 30-second cutoff, Phi-4 can encode long-form audio for speech understanding. For comparability, we also use the same 30-second segments for Phi-4-multimodal-instruct. For readability, we will refer to this model as phi4-multimodal in the remainder of the paper.
\end{itemize}

\textbf{Baselines.} \enspace
For our baselines, we follow the state-of-the-art DL-based method in \cite{amiri_suppress}. To this end, we first obtain feature representations from a foundation model and average over the temporal dimension. We then pass these representations through two linear layers: the first layer projects the representation to 256 dimensions, and the second projects it to 2. We also use a ReLU non-linearity between the layers. As foundation models we consider both wav2vec 2.0 \cite{w2v2} and XLS-R \cite{xlsr}. To choose the layer for obtaining the representations, we follow \cite{tilak} and select the 10th and 16th layers for wav2vec 2.0 and XLS-R, respectively.

\textbf{Training Setup.}
For fine-tuning the LALMs, we used SFT for 5 epochs with a learning rate of \(1 \times 10^{-4}\) for Qwen2-Audio and \(5 \times 10^{-5}\) for Phi-4-multimodal, respectively. We used a cosine learning-rate scheduler. For the Qwen2-Audio model, we added a LoRA \cite{hu2022lora} adapter with $r=64$ and $\alpha= 128$ on all linear layers and kept everything else frozen. For Phi-4-multimodal-instruct, we continued fine-tuning its own speech LoRA adapter and unfroze the audio encoder and projection layer. 

To train the baseline models, we follow the same setting as \cite{amiri2024adversarial}. We use cross-entropy loss for 100 epochs and the Adam optimizer with a learning rate of $1 \times 10^{-4}$ and the \textit{ReduceLROnPlateau} learning rate scheduler with a patience of 5 and a factor of 0.5. We stop training when either the learning rate drops to $1 \times 10^{-4}$ of its original value or when we reach the maximum number of epochs.

For all experiments, we evaluated the model on the validation set at the end of each epoch, and the best model was selected based on the validation set loss. We report speaker-level accuracy across all experiments, for which we perform voting among all utterances belonging to a given speaker to obtain the diagnosis for that speaker.

\textbf{Low-level Acoustic Features.}
As described in Section~\ref{sec:proposed}, we augment the input prompt with low-level acoustic features computed using openSMILE \cite{eyben2010opensmile}. For our experiments, we use the mean and standard deviation of pitch (Hz) and loudness (dB), as well as the mean harmonic-to-noise ratio (dB) and mean speech rate (syllables per second). 
Unlike \cite{mai2025aa}, we use the exact numerical values of these features directly in the input prompt.

\section{Experimental Results}
\label{sec:exp_result}

In this section, we first compare our proposed feature-augmented LALMs against state-of-the-art DL-based baselines and then conduct two ablation studies. The first ablation isolates the contribution of fine-tuning and feature augmentation, comparing zero-shot performance, fine-tuning without additional features, and our proposed fine-tuning with feature augmentation. The second ablation evaluates the performance of our framework as the number of features used for augmentation increases from one to four.

{\bf{Comparison with state-of-the-art baselines.}} \enspace
Table~\ref{tab:baseline} presents results for the state-of-the-art DL-based baselines and for the LALMs fine-tuned with our proposed feature-augmented framework. While the DL-based baselines already achieve a high performance on this task, both LALMs achieve  state-of-the-art performance when fine-tuned with our framework. These results demonstrate the potential of LALMs for dysarthric speech detection. We investigate the source of this improvement in the ablation studies that follow.

\begin{table}[t!]
  \caption{Speaker-level accuracy of the considered approaches on PC-GITA using 10-fold cross-validation.}
  \label{tab:baseline}
  \centering
    \begin{tabularx}{\linewidth}{X|c}
    \toprule
    \multicolumn{1}{X|}{\textbf{Approach}} & \textbf{Accuracy}\\
    \midrule
    \multicolumn{1}{X|}{wav2vec2-based}  & $82.3 \pm 3.1 \% $\\ 
    \multicolumn{1}{X|}{XLS-R-based}  & $80.3 \pm 1.7 \%$\\
    \midrule
    \multirow{1}{*}{Phi-4-Multimodal} (Feature-augmented)& $\underline{83.0 \pm 2.2\%}$\\ 

    \multirow{1}{*}{Qwen2-Audio (Feature-augmented)} & $\mathbf{85.0 \pm 0.8 \%}$\\
    
    \bottomrule
  \end{tabularx}
\end{table}

{\bf{Ablation study: fine-tuning and feature augmentation.}} \enspace
We conduct an ablation study to investigate the zero-shot performance of the LALMs and the effect of fine-tuning with and without our proposed features. Results are provided in Table~\ref{tab:abb}, using the standard prompt described in Table~\ref{tab:prompt} for the zero-shot and standard fine-tuning settings. Both LALMs perform at chance level in the zero-shot setting, indicating that they do not inherently possess the capability to perform this task without fine-tuning. Fine-tuning with the standard prompt achieves competitive performance compared with the DL-based baselines, though it still lags behind our feature-augmented setting. We should note that feature augmentation also reduces the standard deviation of performance across seeds, suggesting that it makes the models more robust to how folds are constructed.
Taken together, these results show that fine-tuning accounts for the majority of the improvement over zero-shot performance, with feature augmentation providing a smaller, but consistent, additional gain.

\begin{figure}[h!]
  \centering
  \includegraphics[width=0.9\linewidth]{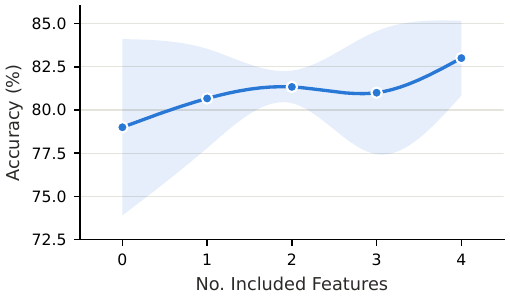}
  \vspace{-0.3cm}
  \caption{The effect of incremental feature addition.}
  \label{fig:abb}
  \vspace{-0.3cm}
\end{figure}

{\textbf{Ablation study: individual acoustic features.}} \enspace
To assess the effect of augmenting with an incremental number of features, we conduct a further ablation study using the Phi-4-Multimodal model as a representative example. We first fine-tune the model using each feature individually to rank the four features from most to least effective, yielding the rank: HNR, speech rate, loudness, and pitch. We then incrementally combine features in this order, evaluating HNR alone, HNR and speech rate, HNR, speech rate, and loudness, and finally all four features together. The task description remains general and is not modified based on which features are introduced. Results are shown in Fig.~\ref{fig:abb}, reporting mean and standard deviation for the standard formulation without feature augmentation and for the proposed framework under each incremental-combination setting. It can be observed that incrementally adding features leads to an overall improvement in model performance, with all four features combined achieving the best result.
\begin{table}[t!]
\caption{Ablation study of zero-shot performance and the feature-augmented framework.
}
\label{tab:abb}
  \centering
\begin{tabularx}{\linewidth}{X|ccccc}
\toprule

\multirow{1}{*}{Model} &
\multicolumn{2}{c}{Feature Mode} & \multicolumn{2}{|c}{Accuracy} \\


\midrule

\multirow{3}{*}{Phi-4-Multimodal} 
& \multicolumn{2}{c}{Zero-shot} & \multicolumn{2}{|c}{$50.0 \pm 0.0\%$}  \\
& \multicolumn{2}{c}{Standard} & \multicolumn{2}{|c}{$79.0 \pm 5.1\%$}  \\
& \multicolumn{2}{c}{Feature-augmented} & \multicolumn{2}{|c}{$\mathbf{83.0 \pm 2.2\%}$}  \\

\hline

\multirow{3}{*}{Qwen2-Audio} 
& \multicolumn{2}{c}{Zero-shot} & \multicolumn{2}{|c}{$51.0 \pm 0.0\%$}  \\
& \multicolumn{2}{c}{Standard} & \multicolumn{2}{|c}{$84.0 \pm 2.5\%$}  \\
& \multicolumn{2}{c}{Feature-augmented} & \multicolumn{2}{|c}{$\mathbf{85.0 \pm 0.8\%}$}  \\


\bottomrule
\end{tabularx}
\vspace{-0.3cm}

\end{table}

\section{Conclusion}
In this work, we proposed a framework for adapting LALMs to dysarthric speech detection by augmenting the textual input with low-level acoustic features and speaker demographic information. Our experiments showed that this framework consistently improves the performance of LALMs. We also conducted ablation studies to investigate the initial zero-shot performance of the models and the effect of adding features to the prompt. Our findings demonstrate the potential of LALMs for dysarthric speech detection and motivate further research in this area.

{
\small
\bibliographystyle{IEEEbib}
\bibliography{strings,refs}
}

\end{document}